\documentclass[aps,twocolumn,epsfig,rotate,showpacs]{revtex4}
\usepackage{epsfig}
\usepackage{graphicx}
\usepackage{amsmath}

\newcommand{\beq}{\begin{eqnarray}}
\newcommand{\eeq}{\end{eqnarray}}
\begin{document}       
\title{Generic Quantum Ratchet Accelerator with Full Classical Chaos}
\author{Jiangbin Gong and Paul Brumer}
\affiliation{Chemical Physics Theory Group, Department of Chemistry, and
Centre for Quantum Information and Quantum Control\\
University of Toronto, Toronto, M5S 3H6, Canada}
\date{\today}
\begin{abstract}
A simple model of quantum ratchet transport
that can generate
unbounded linear acceleration
of the quantum ratchet current is proposed, with the underlying classical
dynamics fully chaotic.
The results demonstrate that
generic acceleration of quantum ratchet transport can occur
with any type of classical phase space structure.
The quantum ratchet transport with full classical chaos
is also shown to be very robust to noise due to
the large linear acceleration afforded by the quantum dynamics.
One possible experiment allowing observation of these predictions 
is suggested.
\end{abstract}
\maketitle

Understanding and exploring aspects of quantum transport is of great
importance in a variety of contexts, including, for example,
nanoscale electronic devices \cite{hanggi}, 
atom optics with optical lattices \cite{revmod}, and coherently controlled 
photocurrents in semiconductors
\cite{brumerbook}.
Of particular interest is
directed transport without a biased field \cite{Reimann02,Hanggitoday},
an important phenomenon often called ratchet transport.
While ultimately both quantum and classical ratchet 
transport are of similar origin, namely, breaking of
certain spatial-temporal symmetries,
it is challenging to understand their relationship in general \cite{brumer06}.
Indeed, as suggested by
quantum tunneling induced current reversal
\cite{Reimann,Lehmann,Linke,gongpre04},
knowledge of classical transport
should be applied with caution in understanding quantum transport.

The quantum-classical contrast in ratchet transport is all the more complex
and intriguing in chaotic systems, where the symmetry-breaking that underlies
directed quantum transport 
is necessarily entangled with the many aspects of quantum chaos, {\it e.g.},
dynamical localization, quantum resonance, and quantum anomalous diffusion.
As such, chaotic model systems are especially attractive 
for studies of quantum ratchet transport, both theoretically and experimentally
\cite{Schanz,Monteiro,Jones,gongpre04,Casati,Lundh,libw}.
Consider the familiar kicked-rotor model and its extensions as examples.
It has been shown that dynamical localization 
can saturate quantum directed current \cite{gongpre04}, and that
quantum resonance can induce linear acceleration of the directed current
without saturation, even when the underlying classical dynamics is fully
chaotic \cite{Lundh,libw}.
The latter result is quite surprising, insofar as the ensemble-averaged
classical acceleration rate of directed current
should, according to a recent classical theorem \cite{Schanz}, 
vanish when the classical dynamics is completely chaotic.  
However, 
this type of directed quantum transport occurs only for isolated 
values of the effective Planck constant and is extremely
vulnerable to noise \cite{Lundh}. Hence 
an experimental observation of quantum resonance induced
ratchet transport is not expected in the near future.
Indeed, solely observing the 
ballistic transport associated with quantum resonance, not to mention
directed transport, is already a highly demanding atom optics experiment
\cite{phillips}.

In this paper we propose a novel
quantum ratchet transport model, called a 
``quantum ratchet accelerator", that also displays unbounded linear acceleration
of the quantum ratchet current while the underlying
classical dynamics is fully chaotic. However,
this quantum ratchet accelerator is {\it generic}:
it is unrelated to quantum resonance, and the
acceleration rate is in general nonzero for an
arbitrary effective Planck constant. 
This firmly establishes, for the first time, that
generic quantum ratchet transport as well as its acceleration
is possible with full chaos
in the underlying classical dynamics, and hence does not
require mixed classical phase space structures.
Equally important,  
quantum ratchet transport in our quantum ratchet accelerator
is robust to noise effects. As shown below,
noise of considerable intensity will
saturate the otherwise accelerating ratchet current, but 
large ratchet currents that are absent in the
classical system are still observed.
One possible experimental scenario to observe the proposed acceleration
is also suggested below.

The model proposed below is an extension of the 
kicked Harper model, a paradigm of quantum chaos that 
may not display dynamical localization
\cite{Geisel,Casati92,Lima}. The kicked 
Harper model has attracted enormous interest
and is closely related to driven harmonic 
oscillator systems \cite{Zaslavskybook}, kicked
charges in a magnetic field \cite{Danapla}, 
and driven electrons on the Fermi surface 
\cite{Fishman}.  Here we re-approach this paradigm from the perspective
of quantum ratchet transport, by breaking the spatial symmetry of
the kicking potential. Specifically, 
we consider the following ``delta-kicked" model, 
\begin{eqnarray}
H & = & L\cos(p)+KV(q)\sum_{n}\delta (t-n); \\
\label{Hami}
V(q)& = & \cos(q+\phi_{1})+\eta \sin(2 q+\phi_{2}).
\end{eqnarray}
Here all variables should be understood as appropriately scaled and hence
dimensionless. In particular, 
$q$ and $p$ are position and momentum variables, $n$ is
an integer, and $L$, $K$, $\phi_{1}$, $\phi_{2}$, $\eta$ 
are system parameters. The commutation relation 
between $q$ and $p$ gives the effective Planck constant $\hbar$, {\it i.e.,} 
$i\hbar=[q,p]$. The model reduces to the original kicked Harper model
for $\phi_{1}=0$ and $\eta=0$, and hence 
can be regarded as a generalized kicked 
Harper model. For future theory development we note that Dana \cite{Danapre}
has studied some basic properties of the quasi-energy bands associated with
any function $V(q)$ that has a period of $2\pi$. 

In presenting detailed results we focus mainly
on one typical case, where $\phi_{1}=\phi_{2}=0$, $\eta=1.0$, and $K=2L=3.0$.
Related cases will also be emphasized
when the comparison becomes enlightening.
The phase space structure for the associated classical map in a unit cell
is shown in Fig. 1a, where no stable islands are found on a very fine scale.
Hence, for all practical purposes, the classical
dynamics is fully chaotic. Indeed, 
the broken spatial symmetry, as evident in the kicking potential,
is invisible in Fig. 1a. By contrast, for smaller values of $K$ and $L$,
the phase space becomes structured;
a mixture of chaotic and integrable motions for some parameters (Fig. 1b),
or predominantly integrable (Fig. 1c). Concerning the results in
Fig. 1b and Fig. 1c, we also note (i) that
the spatial symmetry of the phase space structure is clearly broken  
and (ii) that some phase space invariant curves
are extended in momentum space, hence allowing unbounded
acceleration of classical trajectories. The intuitive character of this
acceleration, coupled to the slowness of the $q$-averaged acceleration
rate of the directed currents (described below), make it far
less interesting than the chaotic case emphasized in this paper.

\begin{figure}
\begin{center}
\epsfig{file=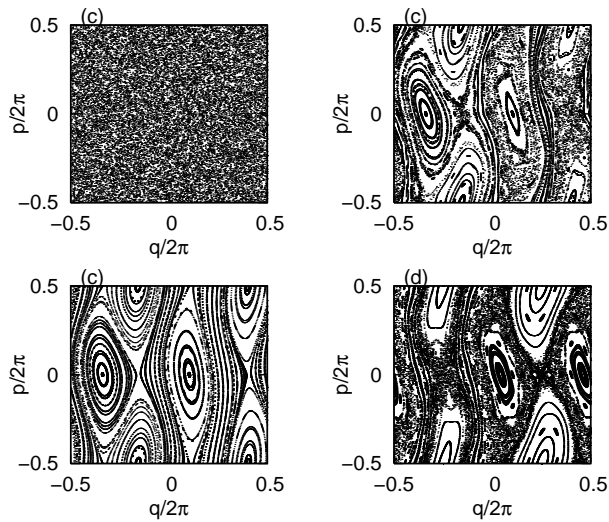,width=8.2cm}
\end{center}
\caption{Classical phase space structures for the
generalized kicked Harper models with 
Hamiltonian given by Eq. (\ref{Hami}).  In (a)-(c) the kicking potential is given by
$V(q)=\cos(q)+\sin(2q)$, and in (d) $V(q)=\sin(q)+\cos(2q)$.  $K=2L=3.0$ in (a),
$K=2L=1.0$ in (b) and (d), and $K=2L=0.4$ in (c). 
As shown in Fig. 2, cases (a)-(c) can be exploited to 
generate quantum ratchet current. Not so for case (d) due to
a spatial-temporal symmetry.}
\end{figure}

Consider now the ensemble-averaged (i.e., averaged over all $q$)
quantum current $\langle p\rangle$, for
an initial ensemble $p=0$ that is symmetric in time and space. To emphasize
that our results are unrelated to quantum resonance, consider
an effective
Planck constant $\hbar=2\pi/(6+\sigma_{g})$, with $\sigma_{g}=(\sqrt{5}-1)/2$.
This ensures that $\hbar/2\pi$ 
is as irrational as possible.  
As seen from Fig. 2, the computed quantum current (solid line) displays
beautiful linear acceleration without 
saturation. Indeed, the absence of dynamical localization in this system
implies that the acceleration should continue, unsaturated, with
additional kicks.  The inset
shows the highly asymmetric
probability density distribution after 1000 kicks,
in terms of the basis states $|m\rangle$, with $p|m\rangle=m\hbar|m\rangle$.
Note that the acceleration rate, defined as the average increase of
quantum ratchet current after each kick, is as large as 0.42, yielding a quantum
current that can be orders of magnitude larger than in previous studies
\cite{Monteiro,gongpre04,Casati}.  
By contrast,
an inspection of the classical transport behavior (dotted line)
shows that there is no systematic acceleration in the classical current. 
Rather, the classical current quickly saturates and remains
extremely small at all times. Figure 2 also shows that the
acceleration rate of the quantum current is much smaller, but still nonzero,
for smaller values of $K$ and $L$ that give rise to
either mixed or predominantly integrable dynamics (see Fig. 1b and Fig. 1c).
This makes it clear that all 
types of classical phase space structures, {\it i.e.},  fully chaotic, mixed, or completely integrable,
can be exploited in constructing
a quantum ratchet accelerator, thereby adding an important feature to current
knowledge regarding  quantum-classical correspondence in
ratchet transport \cite{Monteiro,brumer06,gongpre04,Casati,Schanz,Casati92}.
The full chaos case is our focus 
here because it is counter-intuitive and
can give the largest ratchet current.

\begin{figure}
\begin{center}
\epsfig{file=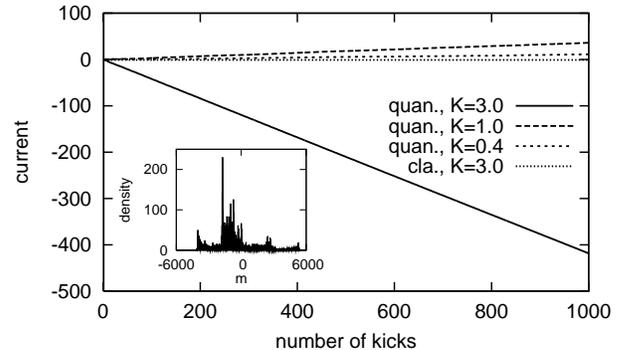,width=8.0cm}
\end{center}
\caption{Unbounded linear acceleration of the 
quantum ratchet current (solid line) with the underlying full classical chaos displayed in
Fig. 1a. The classical ratchet current (dotted line) shows no
linear acceleration,  with its value
indistinguishable from zero on this scale.
The quantum current
associated with mixed ($K=2L=1.0$) or predominantly integrable classical dynamics ($K=2L=0.4$)
is also shown for comparison. The inset shows the highly asymmetric
probability density distribution of the quantum state after 1000 kicks
in representation
of momentum eigenstates, for the fully chaotic case.}
\end{figure}

The sharp contrast between the quantum and classical dynamics 
can be qualitatively understood as follows:
Classically, a system with fully developed chaos will quickly forget 
its history. 
Hence, any classical mechanism related to symmetry breaking can
only operate within the relaxation
time scale, and the ensemble-averaged classical current
should quickly saturate, leading to a vanishing acceleration rate
of the directed current.
However, while all classical phase space
invariant curves get broken in the case of full chaos,
their remnants, typically cantorus-like structures that are much smaller than $\hbar$, can still 
play a key role in the quantum dynamics. In particular,
the cantorus-like structures
present a strong barrier for a time-evolving
quantum wavepacket to pass through, and 
can even attract concentrations of quantum states \cite{Lima}. 
Because these remnants are just as asymmetric as
the classical invariant curves seen in Fig. 1b and Fig. 1c, 
the broken symmetry can be clearly manifested in the quantum dynamics, giving rise to an unbounded linear
acceleration of the quantum current.  Confirming this understanding,
we note that if no classical invariant curves are extended
in the momentum space (e.g., when $2K=L$), then we found no
linear acceleration of the quantum current
for any type of classical phase space structure.

It should also be noted that 
broken spatial symmetry alone does not suffice for ratchet transport.
Consider, for example, a kicking potential $V(q)$ with $\phi_{1}=\pi/2$, 
$\phi_{2}=\pi/2$, and $\eta=1$, {\it i.e.},
$V(q)=\sin(q) + \cos(2q)$. In this case, the spatial symmetry is 
also strongly broken, as clearly seen in Fig. 1d, but both classical
and quantum currents are zero. This
is due to a special temporal-spatial symmetry: the dynamics are
invariant to $q\rightarrow \pi-q$ and $p\rightarrow -p$.
Such a spatial-temporal symmetry can be clearly seen
if we examine either the left or right half of
the phase space cell shown in Fig. 1d.  Averaging over $q$ then causes
loss of directed current.

\begin{figure}
\begin{center}
\epsfig{file=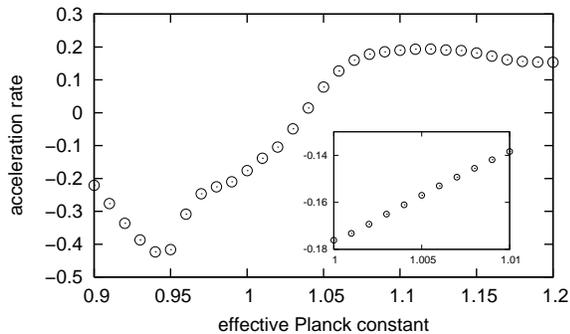,width=7.6cm}
\end{center}
\caption{The $\hbar$-dependence ($0.9\leq \hbar\leq 1.2$)
of the linear acceleration rate of quantum
ratchet current (defined as the average current increase
per kick), with the underlying classical dynamics fully chaotic.
The system parameters are the same as in Fig. 1a.
The smooth $\hbar$-dependence on a smaller scale is shown in the inset.}
\end{figure}

To shed more light on the quantum ratchet accelerator, 
consider now the $\hbar$-dependence of the associated
acceleration rate, for a sampling regime
$0.9\leq \hbar\leq 1.2$ that includes
the $\hbar=2\pi/(6+\sigma_{g})$ case examined in Fig. 2.
Figure 3 displays the results
for 40 values of $\hbar$ ($\hbar=0.9+j/100$, $j=0-30$,
and $\hbar=1.0+m/1000$, $m=1-9$).
The behavior of the acceleration rate
seen in Fig. 3 is highly nontrivial, and 
the quantum current is seen to reverse its sign 
as $\hbar$ increases from 1.03 to 1.04.
This current reversal,
unrelated to quantum tunneling,
reflects a redistribution of
the concentration
of quantum amplitudes on the remnants of classical invariant curves
as $\hbar$ varies.
Further, it suggests that the direction of quantum
ratchet current may be controlled by actively tuning 
the effective Planck constant.
Finally, note that
the inset of Fig. 3 demonstrates that the $\hbar$-dependence
is smooth, suggesting the possibility of a quantitative theory of
quantum ratchet transport.
Such a quantum theory for fully chaotic systems, 
far beyond the scope of this work, would offer a new
tool in understanding quantum chaos.

Since the exposed quantum ratchet transport with full classical chaos
is based entirely on nonclassical effects, one might expect 
it to be fragile when subject to noise.  This is not the case.
As one key advantage of a generic quantum ratchet accelerator,
the quantum ratchet transport has built-in capabilities
to fight against detrimental noise effects.
That is, noise effects
do not easily destroy the ratchet transport, thanks to
the very large acceleration mechanism inherent in the ratchet. 
To see this, consider
one amplitude-noise model as well as one phase-noise model. In the first model,
we assume that the kicking field strength $K$ is scaled by
a randomly fluctuating term $[1+A(\xi-0.5)]$, where $\xi$ is a random
variable uniformly distributed in $[0,1]$.
In the second model, we expand the evolving quantum state in terms of
momentum eigenstates, and then introduce, after each kick,
random phases $\exp[iB2\pi\xi]$ to each individual momentum eigenstate
to dephase them.
The results are shown
in Fig. 4.  Although in both models
higher noise intensity, characterized by larger $A$ or $B$,
suppresses the linear acceleration of quantum ratchet current
to an increasing degree, substantial ratchet currents still survive in the 
presence of noise.
Remarkably, even when the fluctuation in $K$  reaches
10\% of its average value,
or when the random phase fluctuation periodically
introduced to momentum basis states
is  $\approx 0.1\pi$,
the quantum ratchet currents, although saturated, still
remain orders of magnitude larger than
the classical currents and the quantum currents
obtained in, {\it e.g.},  fully chaotic delta-kicked rotor systems
\cite{Monteiro,Casati}.

\begin{figure}
\begin{center}
\epsfig{file=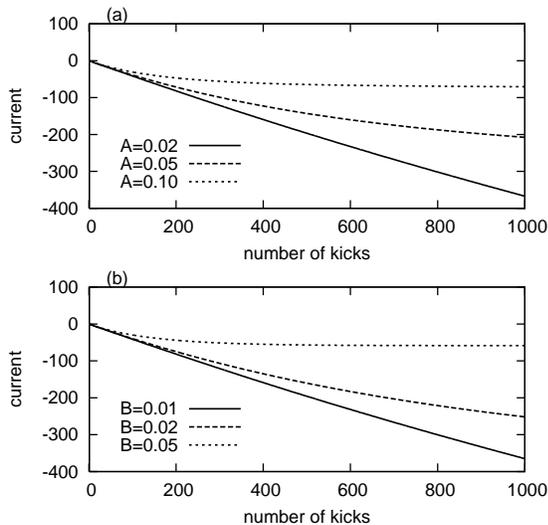,width=7.4cm}
\end{center}
\caption{The robustness of quantum ratchet transport against noise
in a quantum ratchet
accelerator with the underlying full classical chaos shown in Fig. 1a.
The noiseless case is shown as the solid line in Fig. 2.
Effects of amplitude noise
in the kicking field strength $K$ are shown in (a), and effects of
phase noise introduced to momentum eigenstates are shown in (b).
In each case 1000 realizations of noise history are used to obtain
the average behavior.
$A$ and $B$ represent the noise intensity defined in the text.}
\end{figure}

The kicked Harper model is relevant to a number of realistic 
systems \cite{Zaslavskybook,Danapla,Fishman}. As such, the quantum ratchet
accelerator proposed here, a generalized kicked Harper model,
should be of considerable experimental interest.  Consider 
one interesting possibility:
according to a general result of Dana \cite{Danapla},
the time evolution operator of our model can be exactly mapped to 
that of a sub-ensemble of kicked charges in a magnetic field.
The Hamiltonian of the latter is given by
$H=\Pi^2/2 + \lambda U(x,t)\sum_{s}\delta(t-s/4)$, with $\Pi$ the
kinetic momentum of the kicked charge in a magnetic field. The kicking
potential $U(x,t)$ should satisfy, for a particular value of $x_{0}$ 
associated with a constant of the motion that defines the sub-ensemble,
$U(x_{0}-x,t)\propto [L\cos(x)f(t)+ KV(-x)f(t-1/4)+L\cos(x)f(t-1/2)
+KV(x)f(t-3/4)]$, where $f(t)$ satisfies $f(t)=f(t+1)$,
$f(0)=1$, and $f(1/4)=f(1/2)=f(3/4)=0$.  
This type of kicking
potential should be achievable with modern pulse shaping techniques.
Quantum ratchet transport with full classical chaos
would then be observed
in the net charge transport in coordinate space, or
in the asymmetric kinetic momentum distribution of the kicked charges.

Finally, note that the quantum dynamics of
one-dimensional kicked systems
can often be mapped onto that of many-body lattice systems \cite{prosen98}. 
This being the case, our results imply that it is possible for
quantum many-body systems in the thermodynamic limit to
generically display directed 
transport, even when the associated 
classical dynamics is fully chaotic.

In conclusion, we have proposed a generic Hamiltonian
model of quantum ratchet transport where the underlying classical
dynamics is fully chaotic.  Our results add an important feature
to current knowledge of
ratchet transport: generic acceleration of quantum ratchet transport can occur
with any type of classical phase space structure. Further, the exposed
$\hbar$-dependence of
the acceleration rate of the quantum ratchet current represents 
a novel and challenging issue in quantum chaos. Finally, 
the robustness of the quantum ratchet transport against noise 
is one key advantage of this quantum ratchet accelerator.
Future work will also
consider active manipulation of the quantum ratchet transport,
by taking advantage of the $\hbar$-dependence 
of the acceleration rate and by seeking an 
optimized kicking potential to generate ratchet transport most 
efficiently.

{\bf Acknowledgment}:
This work was supported by the Natural Sciences and Engineering
Research Council of Canada.


\begin{thebibliography}{100}
\bibitem{hanggi}S. Kohler, J. Lehmann, and P. Hanggi,  Phys. Rep. {\bf 406}, 379 (2005).
\bibitem{revmod} O. Morsch and M. Oberthaler, \rmp{\bf 78}, 179 (2006).
\bibitem{brumerbook}M. Shapiro and P. Brumer, {\it Principles of the Quantum Control of Molecular Processes} (John Wiley, New York, 2003).
\bibitem{Reimann02} P. Reimann, Phys. Rep. {\bf 361}, 57 (2002).
\bibitem{Hanggitoday} R.D. Astumian and P. Hanggi, Phys. Today {\bf 55} (11), 33 (2002).
\bibitem{brumer06}I. Franco and P. Brumer, \prl{\bf 97}, 040402 (2006).
\bibitem{Reimann} P. Reimann, M. Grifoni, and P. Hanggi, \prl{\bf 79}, 10 (1997).
\bibitem{Lehmann} J. Lehmann {\it et al}., \prl{\bf 88}, 228305 (2002).
\bibitem{Linke} H. Linke {\it et al}., Science {\bf 286}, 2314 (1999).
\bibitem{gongpre04} J. Gong and P. Brumer, \pre{\bf 70}, 016202 (2004).
\bibitem{Schanz} H. Schanz, M.F. Otto, R. Ketzmerick, and
T. Dittrich, \prl{\bf 87}, 070601 (2001).
\bibitem{Monteiro}T.S. Monteiro, P.A. Dando, N.A.C. Hutchings, and M.R. Isherwood, \prl{\bf 89}, 194102 (2002).
\bibitem{Jones} P.H. Jones {\it et al.}, quant-ph/0309149.
\bibitem{Casati} G.C. Carlo, G. Benenti, G. Casati, and D.L. Shepelyansky,
\prl{\bf 94}, 164101 (2005).
\bibitem{Lundh} E. Lundh and M. Wallin, \prl{\bf 94}, 110603 (2005).
\bibitem{libw}D. Poletti, G.C. Carlo, and B. Li, quant-ph/0606124.
\bibitem{phillips} C. Ryu, {I\it et al.}, \prl{\bf 96}, 160403 (2006).
\bibitem{Lima}R. Lima and D. Shepelyansky, \prl{\bf 67}, 1377 (1991). 
\bibitem{Geisel}T. Geisel, R. Ketzmerick, and G. Petschel, 
\prl{\bf 67}, 3635 (1991).
\bibitem{Casati92} R. Artuso {\it et al}, \prl{\bf 69} 3302 (1992).
\bibitem{Zaslavskybook}G.M. Zaslavsky, R.Z. Sagdeev, D.A. Usikov, and A.A. Chernikov, {\it Weak Chaos and Quasi-regular Patterns} (Cambridge Univ. Press, Cambridge, 1991).
\bibitem{Danapla}I. Dana, Phys. Lett. A {\bf 197}, 413 (1995).
\bibitem{Fishman} A. Iomin and S. Fishman, \prl{\bf 81}, 1921 (1998).
\bibitem{Danapre}I. Dana, \pre{\bf 52}, 466 (1995).
\bibitem{prosen98} T. Prosen, \prl{\bf 80}, 1808 (1998).


\end{thebibliography}
\end{document}